\documentstyle[twocolumn,aps,prc,epsfig,epsf,here]{revtex}
\begin{document}

\def\piminuspiplus{1.025}
\def\piminuspiplusstaterr{0.006}
\def\piminuspiplussyserr{0.018}
\def\kminuskplus{0.95}
\def\kminuskplusstaterr{0.03}
\def\kminuskplussyserr{0.03}
\def\pbarp{0.73}
\def\pbarpstaterr{0.02}
\def\pbarpsyserr{0.03}

\def\npart{317}
\def\nparterr{10}

\def\abcorrgheisha{5.1}
\def\abcorrfluka{2.2}
\def\abcorr{3.7}
\def\aberr{1.6}


\title{Ratios of charged antiparticles to particles near midrapidity in Au+Au \\
collisions at $\sqrt{s_{_{NN}}} =$ 200 GeV}
\author{
B.B.Back$^1$,
M.D.Baker$^2$,
D.S.Barton$^2$,
R.R.Betts$^6$,
M.Ballintijn$^4$,
A.A.Bickley$^7$,
R.Bindel$^7$,
A.Budzanowski$^3$,
W.Busza$^4$,
A.Carroll$^2$,
M.P.Decowski$^4$,
E.Garcia$^6$,
N.George$^{1,2}$,
K.Gulbrandsen$^4$,
S.Gushue$^2$,
C.Halliwell$^6$,
J.Hamblen$^8$,
G.A.Heintzelman$^2$,
C.Henderson$^4$,
D.J.Hofman$^6$,
R.S.Hollis$^6$,
R.Ho\l y\'{n}ski$^3$,
B.Holzman$^2$,
A.Iordanova$^6$,
E.Johnson$^8$,
J.L.Kane$^4$,
J.Katzy$^{4,6}$,
N.Khan$^8$,
W.Kucewicz$^6$,
P.Kulinich$^4$,
C.M.Kuo$^5$,
W.T.Lin$^5$,
S.Manly$^8$,
D.McLeod$^6$,
J.Micha\l owski$^3$,
A.C.Mignerey$^7$,
R.Nouicer$^6$,
A.Olszewski$^3$,
R.Pak$^2$,
I.C.Park$^8$,
H.Pernegger$^4$,
C.Reed$^4$,
L.P.Remsberg$^2$,
M.Reuter$^6$,
C.Roland$^4$,
G.Roland$^4$,
L.Rosenberg$^4$,
J.Sagerer$^6$,
P.Sarin$^4$,
P.Sawicki$^3$,
W.Skulski$^8$,
S.G.Steadman$^4$,
P.Steinberg$^2$,
G.S.F.Stephans$^4$,
M.Stodulski$^3$,
A.Sukhanov$^2$,
J.-L.Tang$^5$,
R.Teng$^8$,
A.Trzupek$^3$,
C.Vale$^4$,
G.J.van~Nieuwenhuizen$^4$,
R.Verdier$^4$,
B.Wadsworth$^4$,
F.L.H.Wolfs$^8$,
B.Wosiek$^3$,
K.Wo\'{z}niak$^3$,
A.H.Wuosmaa$^1$,
B.Wys\l ouch$^4$\\
\vspace{3mm}
\small
$^1$~Argonne National Laboratory, Argonne, IL 60439-4843, USA\\
$^2$~Brookhaven National Laboratory, Upton, NY 11973-5000, USA\\
$^3$~Institute of Nuclear Physics, Krak\'{o}w, Poland\\
$^4$~Massachusetts Institute of Technology, Cambridge, MA 02139-4307, USA\\
$^5$~National Central University, Chung-Li, Taiwan\\
$^6$~University of Illinois at Chicago, Chicago, IL 60607-7059, USA\\
$^7$~University of Maryland, College Park, MD 20742, USA\\
$^8$~University of Rochester, Rochester, NY 14627, USA\\
}
\maketitle

\begin{abstract}\noindent
The ratios of charged 
antiparticles to particles have been obtained for pions, kaons, and protons
near midrapidity in central Au+Au collisions 
at $\sqrt{s_{_{NN}}} = 200$~GeV. 
Ratios of  
$\langle \pi^- \rangle / \langle \pi^+ \rangle  =  
\piminuspiplus \pm
\piminuspiplusstaterr \mbox{(stat.)} \pm
\piminuspiplussyserr \mbox{(syst.)}$,
$\langle K^- \rangle/ \langle K^+ \rangle  =  
\kminuskplus \pm
\kminuskplusstaterr \mbox{(stat.)} \pm
\kminuskplussyserr \mbox{(syst.)}$, and 
$\langle \bar{p} \rangle / \langle p \rangle  =  
\pbarp \pm
\pbarpstaterr \mbox{(stat.)} \pm
\pbarpsyserr \mbox{(syst.)}$ have been observed.
The $\langle K^- \rangle/ \langle K^+ \rangle$ and 
$\langle \bar{p} \rangle / \langle p \rangle$ ratios are
consistent with a baryochemical potential $\mu_B$ of 27~MeV,
roughly a factor of 2 smaller than in $\sqrt{s_{_{NN}}} = 130$~GeV collisions.
The data are compared to results from lower energies and model calculations.
Our accurate measurements of the particle ratios impose stringent constraints
on current and future models dealing with baryon production and transport.
\end{abstract}


In this paper, the ratios of antiparticles to particles for
primary charged pions, kaons, and protons in central Au+Au collisions
at $\sqrt{s_{_{NN}}} = 200$~GeV were determined using data from
the PHOBOS detector. The data were taken during the 2001 run of
the Relativistic Heavy-Ion Collider (RHIC) at Brookhaven
National Laboratory. The experiments at the RHIC aim at
understanding the behavior of strongly interacting matter at
high temperature and density, testing predictions of quantum
chromodynamics (QCD).

Earlier studies showed that the $\langle \bar{p} \rangle / \langle p \rangle$
ratio increases from 0.1 observed in fixed-target experiments at $\sqrt{s_{_{NN}}} = 17.3$~GeV\cite{na44,na49}
at the SPS to a value of $0.60 \pm 0.04 \mbox{(stat.)} \pm 0.06 \mbox{(syst.)}$ observed in
PHOBOS at $\sqrt{s_{_{NN}}} = 130$~GeV\cite{phobos130} which is in agreement with the values observed by
the other RHIC experiments\cite{brahms130,phenix130,star130}.
Nucleus + nucleus (A+A) collisions at the RHIC have reached a regime where
the fraction of baryons at midrapidity is not dominated by baryon transport alone.
Therefore, these collisions are approaching the 
net baryon-free regime for which experimental results can be directly
compared to first-principle QCD calculations.
The particle ratios at 130~GeV were shown to be 
consistent with statistical models of particle production \cite{magestro}, 
assuming chemical equilibrium. These models gave a very good description of 
the particle ratios with an assumed hadronization temperature of 
175 MeV and a baryochemical potential $\mu_B = 45 \pm 5$~MeV. However, it proved
difficult to explain the results in terms of microscopic descriptions of the evolution of the particle source
using hadronic transport models \cite{phobos130,hijing,rqmd}.  

The data in this paper allow further tests of models of baryon production and baryon number
transport in nuclear collisions, using a study of the energy dependence of 
particle ratios between $\sqrt{s_{_{NN}}} = 130$~GeV and $\sqrt{s_{_{NN}}} = 200$~GeV.
The results were obtained using the PHOBOS
two-arm magnetic spectrometer.  Details of the setup 
have been previously described\cite{phobos1,phobos2,pak_qm2001}. 
Each arm has a total of 16 layers of silicon sensors (corresponding
to 50000 pixels), providing tracking both outside and 
inside the 2-T field of the PHOBOS magnet. For this analysis, only the 
central region of each arm was used, covering $\pm 15^\circ$ around 
the $45^\circ$ axis of the experiment in the horizontal plane. As three
silicon layers do not extend into this geometrical acceptance region,
particles within this acceptance region typically traversed 13 silicon layers.
A two layer silicon detector (VTX) covering $|\eta| < 1.5$ and 25\% of 
the azimuthal angle provided additional information on the position
of the primary collision vertex. The two arms and frequent magnetic field
polarity reversals allowed for many independent checks in the determination
of particle ratios. The use of highly segmented silicon detectors, beginning
at 10~cm from
the interaction point, result in systematic errors in the correction
to the ratios from secondaries and hyperon decays of 2\% or less.

The primary event trigger was provided by two sets of 16 scintillator 
paddle counters, which covered a pseudorapidity range $3 < |\eta| < 4.5$.
Additional information for event selection was obtained from two 
zero-degree calorimeters which measured spectator neutrons\cite{zdc_nim}.
Details of the event selection and centrality determination 
can be found elsewhere \cite{phobosprl,katzy_qm2001}. 
Monte Carlo (MC) simulations of the apparatus
were based on the HIJING event generator \cite{hijing} 
and the GEANT~3.21 simulation package, folding in
the signal response for scintillator counters and silicon sensors.

For this analysis, events with the 12\%
highest cross section were selected.
A Glauber calculation was used to relate
the fraction of the cross section to
the number of participating
nucleons, $N_{part}$\cite{phobosprl,katzy_qm2001}.
The average number of participants for these events 
was estimated to be $\langle N_{part} \rangle = \npart \pm \nparterr \mbox{(syst.)}$.
The data used in this analysis was taken during the first two
months of RHIC 2001 running and represents only 10\% of the total data set.

\begin{figure}[t]
\centerline{
\epsfig{file=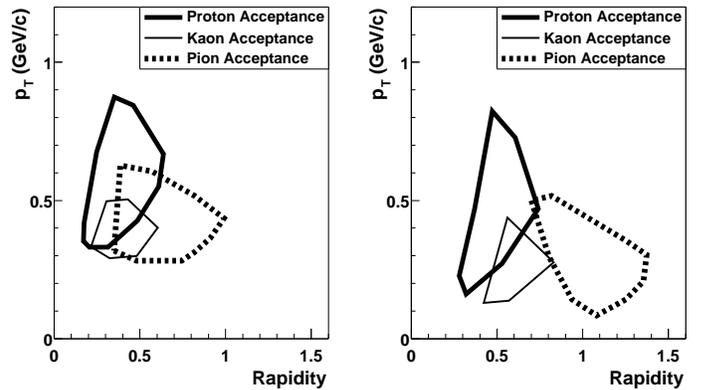,width=9.5cm}}
\caption{Contours of the acceptance of the spectrometer as a function of transverse
momentum and rapidity for pions, kaons, and protons, where the acceptance has fallen
to 10\% of its maximal value. The left plot is for particles bending
towards the beampipe ($\rm{h^+}$,$\rm{B^+}$ or $\rm{h^-}$,$\rm{B^-}$) and the right plot
is for particles bending away from the beampipe ($\rm{h^+}$,$\rm{B^-}$ or $\rm{h^-}$,$\rm{B^+}$).
The acceptance is averaged over the selected vertex range and
the accepted azimuthal angle $\phi$. This results in $p_{T}$ acceptance
ranges of 0.1-0.6~GeV/c for $\pi^{+,-}$, 0.1-0.5~GeV/c for $K^{+,-}$ and 
0.15-0.9~GeV/c for $p,\bar{p}$.}
\label{accept}
\end{figure}

The geometrical layout of the PHOBOS detector led to an asymmetry 
in the acceptance and detection efficiency for positively and 
negatively charged particles, $\rm{h^+}$ and $\rm{h^-}$, at a given magnet polarity,
$\rm{B^+}$ or $\rm{B^-}$ (see Fig.~\ref{accept}). 
Particle and antiparticle yields obtained with opposite field
polarities were used to calculate the
$\langle \pi^- \rangle / \langle \pi^+ \rangle$,
$\langle K^- \rangle/ \langle K^+ \rangle$, and 
$\langle \bar{p} \rangle / \langle p \rangle$ ratios as in
Eqs.~(\ref{ratio_equations}).
\begin{equation}
\begin{array}{ccccc}
\frac{\langle \pi^{-} \rangle}{\langle \pi^{+} \rangle} & : & \frac{N^{B^+}_{\pi^{-}}}{N^{B^+}_{events}}/\frac{N^{B^-}_{\pi^{+}}}{N^{B^-}_{events}} & , & \frac{N^{B^-}_{\pi^{-}}}{N^{B^-}_{events}}/\frac{N^{B^+}_{\pi^{+}}}{N^{B^+}_{events}} \\
&&&&\\
\frac{\langle K^{-} \rangle}{\langle K^{+} \rangle} & : & \frac{N^{B^+}_{K^{-}}}{N^{B^+}_{events}}/\frac{N^{B^-}_{K^{+}}}{N^{B^-}_{events}} & , & \frac{N^{B^-}_{K^{-}}}{N^{B^-}_{events}}/\frac{N^{B^+}_{K^{+}}}{N^{B^+}_{events}} \\
&&&&\\
\frac{\langle \bar{p} \rangle}{\langle p \rangle} & : & \frac{N^{B^+}_{\bar{p}}}{N^{B^+}_{events}}/\frac{N^{B^-}_{p}}{N^{B^-}_{events}} & , & \frac{N^{B^-}_{\bar{p}}}{N^{B^-}_{events}}/\frac{N^{B^+}_{p}}{N^{B^+}_{events}}
\end{array}
\label{ratio_equations}
\end{equation}
It is assumed in the analysis (and checks were performed to verify
this) that the acceptance and efficiency for a specific particle type at
$\rm{B^+}$ is identical to the acceptance and reconstruction efficiency for the
antiparticles of that type at $\rm{B^-}$, and vice versa. This measurement requires
the further assumption that the $p_T$ dependence of the particle and
antiparticle yields within our acceptance is similar. 

The reproducibility of the absolute field strength was found to be 
better than 1\%, based on Hall probe measurements for each polarity
and the comparison of mass distributions for identified particles
for the two polarities.
The trigger event selection was checked using 
the number of reconstructed straight line tracks 
in the low magnetic field region. 
These numbers agreed to within 0.35\% for the $\rm{B^+}$ and $\rm{B^-}$
data sets. An asymmetry in either
the total field strength or the trigger selection would shift
one of the measurements of each ratio in Eqs.~(\ref{ratio_equations})
up while the other would shift down.
Since each shift is equal (and small), this
divergence between the two measurements for each particle type
in Eqs.~(\ref{ratio_equations}) cancels in the average.

Only events with a reconstructed primary vertex position
between -6~cm $< z_{vtx} <$ 6~cm along the beam axis were
included in this analysis. This vertex selection ensures that
both particles and antiparticles can be tracked and identified
in the spectrometer for both bending directions, and thus reduces
vertex-position-dependent systematic effects.
The vertex position was determined by two methods. One
method combined information from both spectrometer arms and
the vertex detector (``global vertex'') and the other only
used information from the arm for which the ratios were 
being measured (``local vertex''). The global vertex is
more precise, but can be affected by misalignments between the
detectors while the local vertex is less precise, but
insensitive to misalignment between the different detectors.

\vspace{-0.5cm}
\vspace{-0.25cm}

The steps of the track reconstruction algorithm have been previously
described\cite{phobos130}. For the present analysis, an 
additional track fitting step was added, which takes into account the
correlations between hit positions along the track. The final track selection was 
based on the $\chi^2$ probability of this fit, improving the rejection
of tracks with incorrectly assigned hits and thereby improving momentum and 
particle identification resolution.

\begin{figure}[t]
\centerline{
\epsfxsize=0.5\textwidth
\epsfbox{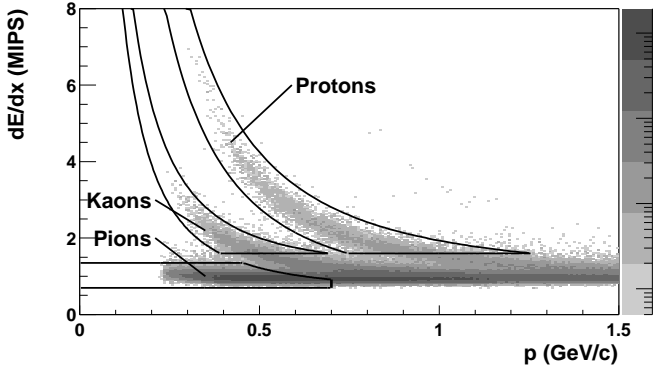}
}
\caption{Distribution of average truncated energy loss as a function of 
reconstructed particle momentum. Three clear bands can be seen,
corresponding to pions, kaons, and protons. The solid lines indicate
the cut regions for counting identified particles.}
\label{dedx_p}
\end{figure}

Particle identification was based on the truncated mean of the specific
ionization $dE/dx$ observed in the silicon detectors. The identification cuts
for pions, kaons, and protons are shown in Fig.~\ref{dedx_p}.
The corresponding acceptance regions for identified particles in 
transverse momentum $p_T$ and 
rapidity are shown in Fig.~\ref{accept}.

The accuracy of the ratio determination relies on the precise cancellation
of acceptance, efficiency, and background effects between the two magnet 
polarity settings.  
To allow a detailed cross-check of this cancellation, 
the data were divided into statistically independent subsets 
(see Table~\ref{table1}). The first division was necessitated by a
significant transverse shift in the beam orbit (0.5~mm) at one point
during data taking. Data before and after this shift
were analyzed separately. The data from the two spectrometer arms were also
analyzed separately. This resulted in four statistically independent subsets. 
These four subsets resulted in eight independent measurements of each 
ratio (each bending direction in the magnet gives a separate measurement).
Each data set was analyzed using both the local and the global 
vertex position, resulting in a total of 16 measurements 
of each ratio. 

\vspace{-0.5cm}

Table~\ref{table1} gives the event and identified particle
statistics for all combinations of magnet polarity, particle charge,
particle species, spectrometer arm and beam orbit condition. The table
contains the results using both the global vertex and
the local vertex as the starting point for the track finding.
The systematic uncertainty due to the vertex finding method was extracted
from the distribution of half of the difference between the value found
using the global vertex and the value using the local vertex. The rms
of this distribution, used as an estimate of the systematic uncertainty,
was found to be $\pm 0.003$ for
$\langle \pi^- \rangle/ \langle \pi^+ \rangle$, $\pm 0.018$ for
$\langle K^- \rangle/ \langle K^+ \rangle$, and $\pm 0.013$ for
$\langle \bar{p} \rangle / \langle p \rangle$.

To perform consistency checks, Eqs.~(\ref{ratio_equations}) were used
to calculate $\langle \pi^- \rangle/ \langle \pi^+ \rangle$,
$\langle K^- \rangle/ \langle K^+ \rangle$, and
$\langle \bar{p} \rangle / \langle p \rangle$ for each of the
eight statistically independent data sets. To check if a systematic
deviation exists between ratios calculated in separate spectrometer
arms, a weighted average of the four ratios calculated in the positive
spectrometer arm was compared against a weighted average of the four
ratios calculated in the negative spectrometer arm. Half of this
difference was taken as an estimate of the systematic error associated
with ratios calculated in separate spectrometer arms. As the pion
ratios have the smallest statistical error, they possess the best
ability to resolve systematic deviations. Therefore, if this difference
for kaons or protons does not exceed a one sigma statistical deviation,
half the pion difference was used. This same method was performed
to estimate systematic errors associated with the other two divisions
of the data. The results of these checks showed that no deviation was
evident beyond the statistical errors for kaons or protons so the pion
ratio's estimate of $\pm 0.015$ for the beam orbit division, $\pm 0.001$
for the arm division, and less than $\pm 0.001$ for the bending direction
division was applied to the kaons and protons.

Table~\ref{table1} also shows the average transverse momentum,
$\langle p_T \rangle$, of identified particles, not corrected
for acceptance and efficiency. For
all particle species, the $\langle p_T \rangle$ of accepted particles
at one magnet polarity agree to within $\pm 3\%$ with the
$\langle p_T \rangle$ of the accepted antiparticles at the opposite
polarity. Also, the average rapidity and the second moments of $p_T$
and $y$ distributions for accepted particles and antiparticles agree
within errors. As a consequence, no acceptance corrections were made
to the particle ratios.

\vspace{-0.25cm}
\begin{table}
\caption{Number of accepted events, accepted identified particles and
average transverse momentum ($\langle p_T \rangle$) in (MeV/c) of
these particles at each magnetic field polarity. Data are
split up by beam orbit condition and spectrometer arm. The number in
parentheses is the number of particles found using the local vertex,
while the number not in parentheses uses the global vertex.
\label{table1}}
\begin{tabular}{|c|rr|cc|rr|cc|}
\multicolumn{8}{|c}{Beam orbit condition 1} & \\
\tableline
 & \multicolumn{3}{c}{$\rm{B^+}$:23725 events} & & \multicolumn{3}{c}{$\rm{B^-}$:20965 events} & \\
\tableline
Narm & \multicolumn{2}{c|}{$N_{particles}$} & $\langle p_T \rangle$ & & \multicolumn{2}{c|}{$N_{particles}$} & $\langle p_T \rangle$ & \\
\tableline
$\pi^+$          & 8532  & (8286)  & $418 \pm 1$ & & 12501 & (12935) & $289 \pm 1$ & \\
$\pi^-$          & 14298 & (14692) & $288 \pm 1$ & & 7643  & (7472)  & $416 \pm 1$ & \\
$K^+$            & 200   & (195)   & $379 \pm 4$ & & 241   & (255)   & $259 \pm 4$ & \\
$K^-$            & 243   & (250)   & $253 \pm 4$ & & 208   & (198)   & $378 \pm 3$ & \\
$p$              & 458   & (472)   & $580 \pm 6$ & & 396   & (405)   & $481 \pm 6$ & \\
$\bar{p}$   & 319   & (324)   & $480 \pm 7$ & & 293   & (291)   & $596 \pm 8$ & \\
\tableline
Parm & \multicolumn{2}{c|}{$N_{particles}$} & $\langle p_T \rangle$ & & \multicolumn{2}{c|}{$N_{particles}$} & $\langle p_T \rangle$ & \\
\tableline
$\pi^+$          & 7139  & (7273)  & $414 \pm 1$ & & 13172 & (13423) & $279 \pm 1$ & \\
$\pi^-$          & 14961 & (15344) & $278 \pm 1$ & & 6432  & (6569)  & $413 \pm 1$ & \\
$K^+$            & 215   & (223)   & $373 \pm 3$ & & 280   & (290)   & $256 \pm 3$ & \\
$K^-$            & 308   & (310)   & $256 \pm 3$ & & 179   & (182)   & $373 \pm 3$ & \\
$p$              & 409   & (427)   & $584 \pm 6$ & & 424   & (433)   & $485 \pm 6$ & \\
$\bar{p}$   & 327   & (346)   & $474 \pm 7$ & & 267   & (264)   & $573 \pm 7$ & \\
\tableline
\tableline
\multicolumn{8}{|c}{Beam orbit condition 2} & \\
\tableline
 & \multicolumn{3}{c}{$\rm{B^+}$:28785 events} & & \multicolumn{3}{c}{$\rm{B^-}$:13824 events} & \\
\tableline
Narm & \multicolumn{2}{c|}{$N_{particles}$} & $\langle p_T \rangle$ & & \multicolumn{2}{c|}{$N_{particles}$} & $\langle p_T \rangle$ & \\
\tableline
$\pi^+$          & 10471 & (10248) & $417 \pm 1$ & & 7407  & (7759)  & $290 \pm 1$ & \\
$\pi^-$          & 16607 & (17267) & $291 \pm 1$ & & 5083  & (4990)  & $417 \pm 1$ & \\
$K^+$            & 304   & (302)   & $376 \pm 3$ & & 137   & (154)   & $257 \pm 5$ & \\
$K^-$            & 278   & (305)   & $257 \pm 3$ & & 135   & (124)   & $373 \pm 4$ & \\
$p$              & 612   & (606)   & $577 \pm 5$ & & 248   & (262)   & $492 \pm 8$ & \\
$\bar{p}$   & 398   & (407)   & $484 \pm 6$ & & 195   & (193)   & $573 \pm 9$ & \\
\tableline
Parm & \multicolumn{2}{c|}{$N_{particles}$} & $\langle p_T \rangle$ & & \multicolumn{2}{c|}{$N_{particles}$} & $\langle p_T \rangle$ & \\
\tableline
$\pi^+$          & 8322  & (8574)  & $414 \pm 1$ & & 8746  & (8946)  & $280 \pm 1$ & \\
$\pi^-$          & 18680 & (19181) & $276 \pm 1$ & & 4229  & (4350)  & $413 \pm 1$ & \\
$K^+$            & 238   & (249)   & $369 \pm 3$ & & 202   & (193)   & $260 \pm 4$ & \\
$K^-$            & 396   & (398)   & $257 \pm 3$ & & 106   & (104)   & $378 \pm 4$ & \\
$p$              & 542   & (565)   & $584 \pm 5$ & & 289   & (298)   & $479 \pm 7$ & \\
$\bar{p}$   & 418   & (434)   & $491 \pm 6$ & & 179   & (197)   & $584 \pm 9$ & \\
\end{tabular}
\end{table}

After the systematic uncertainties of the raw particle ratios were
determined, the weighted average and statistical error for each of
the three ratios was obtained.
The errors associated with vertex finding method and consistency checks
were added in quadrature to give an estimate of the total systematic error.
The result was $\pm 0.015$, $\pm 0.02$, and $\pm 0.02$ for pions, kaons, and protons, respectively.
 
Contamination of the proton and kaon samples by misidentified
particles was checked by testing the stability of the particle ratios
against variations of the particle-ID cuts shown in Fig.~\ref{dedx_p}.
Within the statistical error the ratios were stable against
changes of these cuts. The effect of pion contamination on
$\langle \bar{p} \rangle / \langle p \rangle$ was estimated by
modeling the pion's truncated energy loss distribution as a function of
momentum. A conservative estimate of the number of pions falling within
the proton and antiproton cut region produced a correction of less than
0.8\% to $\langle \bar{p} \rangle / \langle p \rangle$,
which is reflected in its systematic error. This same correction for
$\langle K^{-} \rangle / \langle K^{+} \rangle$ was estimated to be
less than 0.1\%. For pions, electron contamination to
$\langle \pi^{-} \rangle / \langle \pi^{+} \rangle$ was estimated
to be less than 0.2\% based on studies carried out using GEANT
simulations with HIJING events.

To represent the ratios of primary particles produced in the collision,
the detected particle ratios have to be corrected for particles produced
in secondary interactions, loss of particles due to absorption in detector 
material, and feeddown particles from weak decays. Secondary and feeddown
corrections were described in detail in the $\sqrt{s_{_{NN}}} = 130$~GeV
paper \cite{phobos130}. As in that analysis, these
corrections were minimized by requiring tracks to point back to the vertex
within 3.5~mm. The total effect of all corrections to the pion and kaon
ratios was estimated to be less than 1\%; this correction is reflected in
the systematic error.

Corrections to $\langle \bar{p} \rangle / \langle p \rangle$ are more
significant. The secondary correction from the production of protons in the
beampipe and detector material was found to be 0.7\%. This correction is
very small in the spectrometer for the following reasons. Secondary protons
tend to be produced at low momentum. High multiple scattering at low momentum
makes detection efficiency of such secondaries small. Particles are also
required to pass through at least 11 silicon layers. Low momentum
particles tend to bend out of the acceptance (as can be seen from
Table~\ref{table1}, where the $\langle p_T \rangle$ of protons and antiprotons
is above 450~MeV).
The number of secondary particles produced scales
with the number of primary particles produced, mostly pions.
The correction is calculated using HIJING events and, therefore, depends on
how well HIJING reproduces the measured $p/\pi$ ratio and their momentum
distributions. Measurements of $p/\pi$ at 130~GeV\cite{phenix130} are
consistent with the values from HIJING at the same energy. A systematic error
of the size of the correction (1\%) is conservatively assigned.

The absorption correction arises from an asymmetry in the loss of antiprotons versus
protons interacting in the beampipe and planes of the spectrometer. A momentum dependent
correction is used. Only GEANT's ability to model interactions in material plays a role
in this correction. The hadronic interaction packages Gheisha and Fluka were both tested.
Gheisha produced a correction of \abcorrgheisha\% while Fluka produced a correction of
\abcorrfluka\% The correction was therefore estimated from the average as \abcorr\% with a systematic
error contribution of \aberr\% taking into account the errors on each individual package's
result and the difference between those results. This correction was found to be consistent
with analytic calculations.

The feeddown correction is calculated from Eq.~(\ref{feeddown_corr}):
\begin{equation}
\left(\frac{\bar{p}}{p}\left)\left/\left(\frac{\bar{p}+\bar{p}_{_{\bar{\Lambda} \rightarrow \bar{p} \pi^{+}}}}{p+p_{_{\Lambda \rightarrow p \pi^{-}}}}\right)\right.\right.\right. \!\! = \left.\frac{1+br\cdot\varepsilon_{acc}\cdot\frac{\bar{\Lambda}}{\bar{p}}\left(\frac{\bar{p}}{p}\big/\frac{\bar{\Lambda}}{\Lambda}\right)}{1+br\cdot\varepsilon_{acc}\cdot\frac{\bar{\Lambda}}{\bar{p}}}\right.
\label{feeddown_corr}
\end{equation}
This accounts for the difference in the number of hyperons (primarily
$\Lambda$ and $\bar{\Lambda}$) decaying to protons versus antiprotons.
$\varepsilon_{acc}$ is the fraction of protons from lambdas or antiprotons
from antilambdas detected within our acceptance and $br$ is the branching
ratio for $\Lambda \rightarrow p \pi^{-}$ or
$\bar{\Lambda} \rightarrow \bar{p} \pi^{+}$.
$br \cdot \varepsilon_{acc}$ equals 0.3 averaged over $p_T$. The $p_T$
dependence of $\varepsilon_{acc}$ is weak due to two competing effects at
high and low momentum. Higher momentum hyperons decay later due to time
dilation. The decay particles from these hyperons have a reduced chance of
being reconstructed because they are more likely to decay past the first
planes of silicon in our spectrometer. Lower momentum hyperons have decay
particles which deviate more from the original path of their parent and,
therefore, have a lower chance of pointing back to the primary vertex.
The derivation of Eq.~(\ref{feeddown_corr}) assumes that protons and
antiprotons have the same momentum distribution within our acceptance as
lambdas and antilambdas and that the decay proton or antiproton takes most
of the momentum of the lambda or antilambda. The value of
$(\bar{p}/p)/(\bar{\Lambda}/\Lambda)$ was estimated using
a coalescence model to relate this value to $K^-/K^+$, which is measured in
this analysis. Values of $\Lambda/p$, $\bar{\Lambda}/\bar{p}$, and
$K^{-}/K^{+}$ were consistent with coalescence at
130~GeV\cite{phobos130,phenix_lambda}. A $\bar{\Lambda}/\bar{p}$ value of
$0.95 \pm 0.4$, based on the measured value at 130~GeV\cite{phenix_lambda},
was used for the correction. The indicated error in this value includes
the statistical and systematic errors in the measurement at 130 GeV
(0.09 and 0.22, respectively), and an uncertainty associated with
the energy dependence of this ratio. The value of $K^-/K^+$ was allowed
to vary between 0.9 and 1.0 (the total variance allowed by our measurement).
This produced a 1.2\% correction with a systematic error contribution of
2.2\%.

After all corrections we find the following ratios within our acceptance:
\begin{eqnarray*}
&\langle \pi^- \rangle / \langle \pi^+ \rangle & =  \piminuspiplus \pm \piminuspiplusstaterr \mbox{(stat.)} \pm \piminuspiplussyserr \mbox{(syst.)},\\
&\langle K^- \rangle/ \langle K^+ \rangle & =  \kminuskplus \pm \kminuskplusstaterr \mbox{(stat.)} \pm \kminuskplussyserr \mbox{(syst.)},\\
&\langle \bar{p} \rangle / \langle p \rangle & =  \pbarp \pm \pbarpstaterr \mbox{(stat.)} \pm \pbarpsyserr \mbox{(syst.)}.
\end{eqnarray*}

\begin{figure}[t]
\centerline{
\epsfig{file=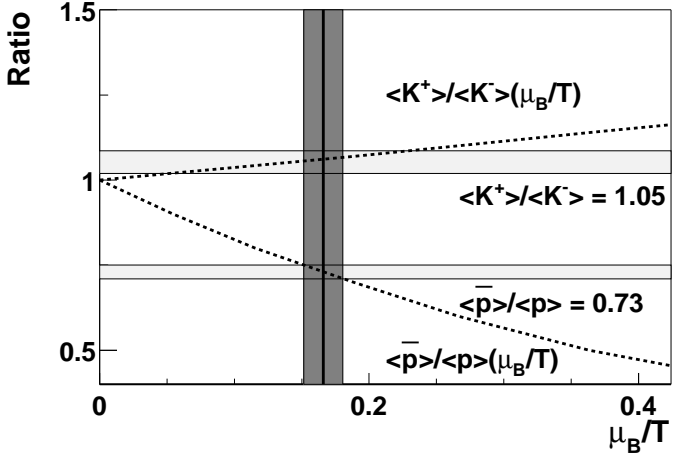,width=9.5cm}}
\caption{Statistical model calculation  (dotted lines) of
$\langle K^+ \rangle/ \langle K^- \rangle$ and
$\langle \bar{p} \rangle / \langle p \rangle$
as a function of $\mu_B/T$ from Becattini \emph{et al.} \protect\cite{redlich_qm2001}.
The horizontal bands show the ratios observed in the data
(statistical errors only). The vertical shaded area indicates
the allowed region in $\mu_B/T$. }
\label{redlich}
\end{figure}

The measured value for $\langle \bar{p} \rangle / \langle p \rangle$
agrees well with the predicted thermal model value of 0.75 at 200 GeV
\cite{thermal_model}. This model assumes baryon number, strangeness,
and charge conservation and, as such, strictly applies to 4$\rm{\pi}$
integrated particle ratios. The prediction is based on a phenomenological
fit to lower energy data, showing that the particle ratios at RHIC follow
the trend observed at lower energies. Using the same statistical model
calculation \cite{redlich_qm2001} as for our 130-GeV data\cite{phobos130},
we estimate $\mu_B/T = 0.17 \pm 0.01 \mbox{(stat.)}$. This estimate is
consistent with the value of $\mu_B/T$ extracted from
$\langle K^+ \rangle/ \langle K^- \rangle$ as shown in Fig.~\ref{redlich}.
For a typical chemical freeze-out temperature of 165 MeV, this estimate
corresponds to $\mu_B = 27 \pm 2 \mbox{(stat.)}$~MeV, which is about half
of the 130-GeV value of $\mu_B = 45 \pm 5 \mbox{(stat.)}$~MeV
\cite{phobos130}.

\begin{figure}[t]
\centerline{
\epsfig{file=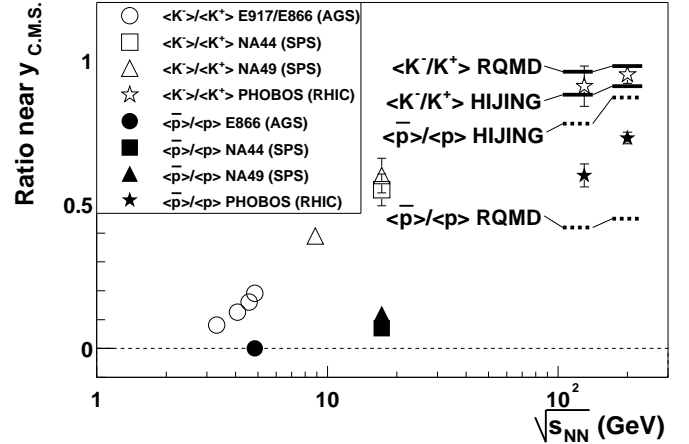,width=9.5cm}}
\caption{$\langle K^- \rangle/ \langle K^+ \rangle$ and
$\langle \bar{p} \rangle / \langle p \rangle$ ratios
as a function of $\sqrt{s_{_{NN}}}$ for nucleus-nucleus collisions,
in comparison with predictions from the HIJING and RQMD models. The
percentage of the highest cross-section used for event selection was
12\% for PHOBOS, 5\% for NA49, 6\% for NA44, 5\% for AGS
$\langle K^{-} \rangle / \langle K^{+} \rangle$, and 8\% for AGS
$\langle \bar{p} \rangle / \langle p \rangle$. Only statistical
errors are shown.}
\label{ratio_roots}
\end{figure}

Our results are compared to lower energy data 
from fixed target experiments\cite{na44,na49,e917} and our 130-GeV
measurement at the RHIC\cite{phobos130} in Fig.~\ref{ratio_roots}. Also shown are
the results of calculations using HIJING\cite{hijing} and RQMD\cite{rqmd}
for 130 and 200 GeV. As was already observed at 130 GeV\cite{phobos130},
both HIJING and RQMD do not agree with the observed value of
$\langle \bar{p} \rangle / \langle p \rangle$. 

In conclusion, the data shown in this paper provide the first information
on the baryon density in central Au+Au collisions at the full RHIC energy.
The observed ratios are in agreement with expectations of statistical 
equilibrium models. A comparison with two hadronic transport codes shows
a significant disagreement with the ratios measured in this analysis. These
ratios provide stringent constraints on current and future models dealing
with baryon production and transport. Future analysis of p+p and d+Au
collisions should provide information on the effect of multiparticle
interactions on particle production.

This work was partially supported by U.S. DOE \linebreak
grants DE-AC02-98CH10886, DE-FG02-93ER40802, \linebreak
DE-FC02-94ER40818, DE-FG02-94ER40865, DE-FG02-99ER41099, and
W-31-109-ENG-38 as well as NSF grants 9603486, 9722606 and 0072204.  The Polish
groups were partially supported by KBN grant 2-PO3B-10323.  The NCU group was
partially supported by NSC of Taiwan under contract NSC 89-2112-M-008-024.

\vspace{-1.0cm}
\begin{thebibliography}{99}
\vspace{-1.5cm}
\bibitem{na44} I.\ G.\ Bearden \emph{et al.}, Phys.\ Lett.\ B\ {\bf 388}, 431 (1996).
\bibitem{na49} J.\ B\"achler \emph{et al.}, Nucl.\ Phys.\ {\bf A661}, 45 (1999).
\bibitem{phobos130} B.\ B.\ Back \emph{et al.}, Phys.\ Rev.\ Lett.\ {\bf 87}, 102301 (2001).
\bibitem{brahms130} I.\ G.\ Bearden \emph{et al.}, Phys.\ Rev.\ Lett.\ {\bf 87}, 112305 (2001).
\bibitem{phenix130} K.\ Adcox \emph{et al.}, preprint, nucl-ex/0112006 (unpublished).
\bibitem{star130} C.\ Adler \emph{et al.}, Phys.\ Rev.\ Lett.\ {\bf 86}, 4778 (2001).
\bibitem{magestro} D.\ Magestro, hep-ph/0112178 (unpublished).
\bibitem{hijing} M.\ Gyulassy and X.\ N.\ Wang, Comput.\ Phys.\ Commun.\ {\bf 83}, 307 (1994). We used HIJING V1.35 with standard parameter settings.
\bibitem{rqmd} H.\ Sorge, Phys.\ Rev.\ C\ {\bf 52}, 3291 (1995). We used version 2.4, including rope formation.
\bibitem{phobos1} B.\ Back \emph{et al.}, Nucl. Phys. {\bf A661}, 690 (1999).
\bibitem{phobos2} H.\ Pernegger \emph{et al.}, Nucl. Instrum. Methods Phys. Res. A {\bf 419}, 549 (1998).
\bibitem{pak_qm2001} B.\ B.\ Back \emph{et al.}, Nucl. Phys. {\bf A698}, 416 (2002).
\bibitem{zdc_nim} C.\ Adler \emph{et al.}, Nucl. Instrum. Methods Phys. Res. A {\bf 461}, 337 (2001).
\bibitem{phobosprl} B.\ B.\ Back \emph{et al.}, Phys.\ Rev.\ Lett.\ {\bf 85}, 3100 (2000).
\bibitem{katzy_qm2001} B.\ B.\ Back \emph{et al.}, Nucl. Phys. {\bf A698}, 555 (2002).
\bibitem{phenix_lambda}  K.\ Adcox \emph{et al.}, preprint, nucl-ex/0204007.
\bibitem{thermal_model} P.\ Braun-Munzinger, D.\ Magestro, K.\ Redlich, and J.\ Stachel, Phys.\ Lett.\ B\ {\bf 518}, 41 (2001).
\bibitem{redlich_qm2001} F.\ Becattini, J.\ Cleymans, A.\ Ker\"anen, E.\ Suhonen, and K.\ Redlich, Phys. Rev. C {\bf 64}, 024901 (2001).
\bibitem{e917} L.\ Ahle \emph{et al.}, Phys.\ Lett.\ B\ {\bf 490}, 53 (2000); \\
 Phys.\ Rev.\ C\ {\bf 60}, 064901 (1999); \\
 Phys.\ Rev.\ Lett.\ {\bf 81}, 2650 (1998).
\end{thebibliography}
\end{document}